\documentclass[12pt]{iopart}
\usepackage{amsfonts}

\usepackage{graphicx}% Include figure files
\usepackage{bm}% bold math

\newcommand{\id}{\mathbb{I}}

\newcommand{\ket}[1]{\mathinner{|{#1}\rangle}}

\newcommand{\ketbra}[2]{|#1\rangle \langle #2 |}

\newcommand{\text}[1]{\textrm{#1}}
\newcommand{\eqref}[1]{Equation (\ref{#1})}

\begin{document}

\title{Overcoming phonon-induced dephasing for indistinguishable photon sources }

\author{Tom Close}
%\email{tom.close@materials.ox.ac.uk}
\address{Department of Materials, Oxford University, Oxford OX1 3PH, UK}

\author{Erik M. Gauger}
\address{Centre for Quantum Technologies, National University of Singapore, 3 Science Drive 2, Singapore 117543}
\address{Department of Materials, Oxford University, Oxford OX1 3PH, UK}

\author{Brendon W. Lovett}
\ead{b.lovett@hw.ac.uk}
\address{SUPA, School of Engineering and Physical Sciences, Heriot Watt University, Edinburgh, EH14 4AS}
\address{Department of Materials, Oxford University, Oxford OX1 3PH, UK}

\begin{abstract}
Reliable single photon sources constitute the basis of schemes for quantum communication and measurement based quantum computing. Solid state single photon sources based on quantum dots are convenient and versatile but the electronic transitions that generate the photons are subject to interactions with lattice vibrations. Using a microscopic model of electron-phonon interactions and a quantum master equation, we here examine phonon-induced decoherence and assess its impact on the rate of production, and indistinguishability, of single photons emitted from an optically driven quantum dot system. We find that, above a certain threshold of desired indistinguishability, it is possible to mitigate the deleterious effects of phonons by exploiting a three-level Raman process for photon production.
\end{abstract}

\maketitle

\section{Introduction}
% [ Use of single photon sources ]

Single photon sources are an essential component of many quantum information processing (QIP) protocols, from quantum key distribution (QKD) protocols \cite{bb84, ekert:91} to linear optical quantum computing (LOQC) schemes \cite{knill:2001, kok07, kok10}.
Optical schemes using path erasure, a two-photon interference effect, can be used to generate long-range entanglement between physically separated systems  \cite{barrett+kok, bose99, simon:03, lim05}. Such procedures can be repeated on many different pairs of systems and so create a distributed cluster state~\cite{raussendorf01,kok10}, which is the key resource required for implementation of measurement based quantum computing.

In order to be useful in these applications, a photon source must be of a high quality in two respects: it must reliably produce a single photon on demand, and the photons produced must be indistinguishable from one another. 

To be perfectly indistinguishable, photons must have the same pulse width, band width, polarization, arrival time at the detector, and carrier frequency. Indistinguishability is vital if photons are to exhibit high quality quantum interference, and its consideration is therefore critical when designing photon sources for LOQC and path erasure entanglement generation.
If photons can be distinguished even in principle, this can lead directly to imperfect LOQC gates, or to a lessening of the degree of entanglement generated in distributed cluster states. In QKD, indistinguishability is less important as interference effects are not required, but it is of paramount importance that no more than one photon is emitted on demand; multiple photon emission leads to security loopholes~\cite{kok10}.

\begin{figure}[htb]
  \begin{center}
  \includegraphics[width=8cm]{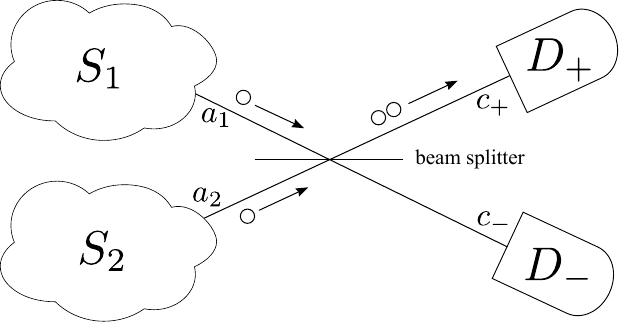}
  \end{center}
  \caption{Hong-Ou-Mandel effect: a pair of photons from two sources $S_1$ and $S_2$ incident on a beam splitter exhibit perfect bunching behaviour if they are completely indistinguishable. In this case the detectors  $D_+$ and $D_-$ will never click simulataneously.}
  \label{HOM_expt}
\end{figure}

As a measure of indistinguishability we exploit the Hong-Ou-Mandel (HOM) effect \cite{HOM}, which relies on the bunching behaviour of identical photons when they are incident on a beam splitter with the same temporal profile. If the photons are indistinguishable, they will always emerge in and be detected in the same output arm (Fig.~\ref{HOM_expt}). The number of same-arm detection events are usually plotted as a function of arrival time of the two photons - and hence a `dip' at a time difference of zero is an indicator of indistinguishability.
The degree of distinguishability is typically given by the visibility of the dip, $v_{\rm HOM}$, which is the proportion of same-arm detections, over many runs of the experiment.

As mentioned above, besides indistinguishability an essential characteristic of a good photon source is that it will consistently produce a single photon, but never more than one, on demand~\cite{lounis05}.  A laser source can be attenuated so that it gives zero or single photons most of the time. However, photons from classical lasers sources obey Poissonian statistics so that in order to keep the two-photon event rate low the probability of a single photon may become unworkably small for many applications. In addition to this, Poissonian sources are unsuitable for two-photon interference experiments since the rate of two photon production from a single source is similar to that for a single photon from each source~\cite{kok10}.
%In contrast, low-dimensional optically active quantum systems such as self-assembled semiconductor quantum dots \textcolor{red}{REFS} do not suffer from this problem and routinely achieve near perfect anti bunching in $g_2$  experiments  \textcolor{red}{REFS}.
It is worth noting that perfect efficiency should never be a requirement in any realistic optical QIP scheme, as these schemes must always be tolerant to photon loss within other parts of the apparatus. However, a good photon source must be reasonably efficient to be useful, and we should not be forced to trade efficiency for other desirable characteristics.

% [ Schemes that have been suggested in the past ]

The use of low-dimensional quantum systems as single photon sources avoids the efficiency problems of Poissonian sources. Successful experimental implementations have been realised in a number of different systems including atom-cavity schemes \cite{rempe:prl:02, kimble:sci:04, grangier:sci:05, rempe:nat-phys:07}, quantum dots \cite{arakawa:nat-mat:06, imamoglu:nat:07} and diamond colour centres \cite{roch:njp:07, bernien:prl:12, lukin:prl:12}. While the majority of the work has been focussed on the efficient production of a single photon, recent experiments in NV centres \cite{bernien:prl:12, lukin:prl:12} and quantum dots~\cite{flagg10, patel10} have demonstrated two-photon interference effects from different sources, albeit sacrificing efficiency by filitering out undesired frequencies. It has been suggested that cavities could be used in these systems, to enhance the emission into the target mode, reducing the need for filtering \cite{greentree:08}.

In order to improve the characteristics of a photon source, it is not sufficient to simply consider the material parameters of the system being used: one should also consider the approach used to control the system. Perhaps the simplest strategy is to excite the system first optically, either coherently or incoherently, and wait for the system to relax into its ground state, emitting a photon in the process; we will henceforth refer to this as the `pulse-relax' technique. % focussing on coherent exitation. 
This approach makes minimal resource demands on the system and, due to its simplicity, is the technique proposed in some remote entanglement generation schemes~\cite{barrett+kok,simon:03}. The pulse-relax approach is problematic in systems where the excited state is sensitive to decoherence, which will degrade the photon's indistinguishability \cite{kiraz:2004, santori:2009, barrett+kok, nazir09}. These effects can be reduced, for example by exploiting the Purcell effect to enhance the emission rate into the desired photon mode~\cite{purcell46,englund05} or by using temporal post-selection of emitted photons~\cite{nazir09}. However with experimental limits on cavity couplings both of these inevitably lead to lower efficiency as the proportion of emissions that are utilized falls \cite{auffeves:prb:10}.

A fundamentally different approach is to use more elaborate QED schemes to release a photon from the system in a controlled manner. In particular single photon sources using a Raman approach have been analysed \cite{kiraz:2004, santori:2009} and experimentally realised \cite{rempe:prl:02}. The approach places more demands on the system, requiring a  three level system with a $\Lambda$-system configuration.
\begin{figure}[htb]
  \begin{center}
  \includegraphics[width=8cm]{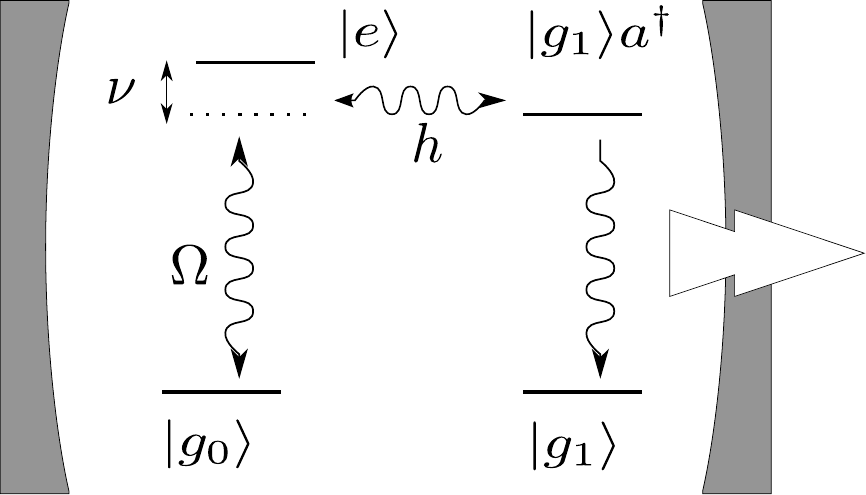}
  \end{center}
  \caption{Effective level structure of a driven $\Lambda-$system inside an optical cavity. Here, $\Omega$ is the amplitude of an external (laser) driving field, $h$ the optical dipole-cavity-coupling, and $\nu$ the shared detuning of the laser and the cavity transition frequency  from the excited state. The system can decay into state $\ket{g_1}$ and in doing so emit a photon into a well-defined external mode.}
  \label{lambda_system}
\end{figure}
Each arm is coupled to either a classical or a quantum light field; in Fig~\ref{lambda_system} we show the situation when one arm is driven classically, with a coupling $\Omega$, and the other arm is coupled to a cavity mode with strength $g$. By detuning both arms of the $\Lambda$-system by the same amount $\nu$, population is transferred from one arm to the other, whilst suppressing population in the top state. Provided that the coupling strengths are small in comparison to the detuning, $\Omega \ll \nu$ and $g\ll\nu$, we induce an effective coupling between the two low-lying states, which causes oscillations with Rabi frequency ${g\Omega}/{4 \nu}$. The population in the excited state remains small at all times, and so any decoherence that arises due to environmental coupling to this excited state may be reduced using this strategy. In particular, a common realisation of a single photon source is a quantum dot, in which the excited state typically has a different charge configuration to the ground state, and so couples to acoustic phonons~\cite{ramsay:2010} that can act as a noise source.
% [ Overview of what we will do ]

In this paper we provide a detailed and realistic analysis of the effects of the lattice vibrations for single photon emitters based on (self-assembled) semiconductor quantum dots. By comparing a standard pulse-relax approach with the aforementioned Raman technique, we find that the latter can can offer vast improvements in terms of both photon indistinguishability and source efficiency. Our results extend and complement a previous study \cite{santori:2009} which considered generic pure dephasing noise. However, here we also show that the precise choice of control parameters is important if phonon-induced decoherence is to be successfully suppressed.

\section{Our model}

% [ The basic setup ]

For convenience of notation we consider the $\Lambda$-system detailed in Fig.~\ref{lambda_system} for both the pulse-relax and the Raman approach. 
%In the former case we are only interested in the transition from one of ground states to the common excited state. 
In both cases one arm of the system is coupled to the cavity with detuning $\nu$. % where we denote the frequency mismatch between cavity and $\Lambda$-system transition by $\nu$. 
In the Raman scheme, the other arm is driven with strength $\Omega$ by a laser with a matching detuning $\nu$, and the system starts in state $\ket{g_0}$. By contrast, we model the pulse-relax approach by setting $\Omega = \nu = 0$ and starting in state $\ket{e}$, ignoring the details of the excitation process. The fact that we are neglecting the excitation step in this simplified picture will slightly favour the pulse-relax approach, but is largely justified on the assumption that the initial excitation process takes place quickly compared with other system dynamics. Our framework thus allows us to consider the pulse-relax approach as a special case of the master equation we will now derive for the Raman approach.

We split the Hamiltonian into contributions from the emitter system and cavity ($sys$), the driving laser ($dr$), the unperturbed phonon bath ($ph$) and a system-phonon interaction term ($int$):
\begin{eqnarray}
  H = H_{sys} + H_{dr} + H_{ph} + H_{int}~.
\end{eqnarray}
In the following, we shall describe each of these parts separately. 

Using the standard Jaynes-Cummings Hamiltonian, the $\Lambda$-system coupled to a cavity mode with strength $h$ and detuning $\nu$ from resonance  can be described by the following Hamiltonian:
\begin{eqnarray}
  H_{sys} = \left(\omega + \nu\right)\ketbra{e}{e} + \omega a^\dagger a + \frac{h}{2}\left(\ketbra{g_1}{e}a^\dagger + \text{h.c.} \right)~,
\end{eqnarray}
where $\text{h.c.}$ denotes the Hermitian conjugate. $\omega$ is the cavity mode frequency.

We can restrict ourselves to one or zero cavity photons since the Jaynes-Cummings model preserves excitation number and, due to the structure of the system, once a photon has escaped re-excitation is impossible. This allows us to replace $\ket{g_1}a^\dagger$ with a new combined atom-photon state $\ket{g_a}$ and $a^\dagger a$ with $\ketbra{g_a}{g_a}$. 

The term $H_{dr}$ describes a laser driving the transition $\ket{g_0} \leftrightarrow \ket{e}$ with coupling strength $\Omega$, and whose frequency is detuned from resonance by the same $\nu$ parameter, yielding
\begin{eqnarray}
  H_{dr} = \Omega \cos(\omega t) \left(\ketbra{e}{g_0} + \text{h.c.} \right)~.
\end{eqnarray}
After performing the rotating wave approximation, assuming that $\omega \gg \nu, h, \Omega$, we are left with 
\begin{eqnarray}
  H_{sys} + H_{dr} =  \nu\ketbra{e}{e} + \frac{1}{2}\left(h \ketbra{g_a}{e} + \Omega \ketbra{e}{g_0}\right) + \text{h.c.}
\end{eqnarray}
for the total system Hamiltonian.

We model the phonons as a bath of harmonic oscillators
\begin{eqnarray}
  H_{ph} &= \sum_q \omega_q b_q^\dagger b_q,
\end{eqnarray}
which couple to the exciton state $\ket{e}$ through the deformation coupling with coupling  constants $f_q = D|q|$ in the usual way \cite{mahan00}: 
\begin{eqnarray}
  H_{int} = \ketbra{e}{e}\sum_q f_q \left(b_q^\dagger + b_q\right).
\end{eqnarray}

The effect of the phonon bath on a driven quantum dot can be described using a Lindblad master equation \cite{gauger:2008}, where the Lindblad operators  induce phonon-assisted transitions between the dressed system eigenstates \cite{gauger:2010}. Being analogous to Refs \cite{gauger:2008, gauger:2010}, it suffices to outline the derivation of the  phonon master equation only briefly in the following. To derive the master equation we must first diagonalise the system Hamiltonian, $H_{sys} + H_{dr}$. We find that the eigenvalues are
\begin{eqnarray}
 \lambda_0 &=& 0 \nonumber, \\
 \lambda_\pm &=& \frac{\nu \pm \sqrt{\nu^2 + \Omega^2 + h^2}}{2},
\end{eqnarray}
with corresponding eigenvectors
\begin{eqnarray}
  \ket{\psi_0} &=& n_0 \left(h\ket{g_0} - \Omega\ket{g_a}\right), \nonumber \\
  \ket{\psi_\pm} &=& n_\pm \left( \Omega\ket{g_0} + h\ket{g_a} +2\lambda_\pm \ket{e} \right).
\end{eqnarray}
$n_0$ and $n_\pm$ are appropriate normalisation factors. Making the Born and Markov approximations leads to the master equation \cite{breuer02} in standard Lindblad form:
\begin{eqnarray}
  \dot{\rho} &= i\left[ \rho, H \right] + D_{ph}(\rho)
\end{eqnarray}
with the phonon dissipator given by
\begin{eqnarray}
  D_{ph}(\rho)  &= J(\Lambda) \left[ \left( N(\Lambda)+1 \right)D\left[ P_\Lambda \right] \rho + N(\Lambda)D[P_\Lambda^\dagger] \rho \right] 
  \label{eqn:phonondiss}
\end{eqnarray}
where $D[L]\rho = L\rho L^\dagger - 1/2(L^\dagger L\rho + \rho L^\dagger L)$, $\Lambda = \lambda_+ - \lambda_- $, and $P_\Lambda = -\ketbra{\psi_-}{\psi_+}$. Note that the phonons only induce transitions between the two optically bright system eigenstates and do not couple to the dark $\ket{\psi_0}$. In the above equation $N(\Lambda)$ is the bosonic mode occupation number:
\begin{eqnarray}
  N(\omega) &= \frac{1}{e^{\beta \omega}-1}
\end{eqnarray}
$\beta = (k_BT)^{-1}$ and we shall be henceforth consider all systems at room temperature, $T = 298$~K. The spectral density function $J(\omega)$ represents the electron-phonon coupling weighted by the density of phonon modes~\cite{breuer02}. We expect this to be dominated by deformation potential coupling, and in this case we obtain~\cite{gauger:2008}:
\begin{eqnarray}
  J(\omega) &= \alpha \omega^3 e^{-\left(\frac{\omega}{\omega_c}\right)^2}~.
  \label{eq:specdens}
\end{eqnarray}
We take $\alpha = 0.0027$~ps$^{-1}$ and $\omega_c = 2.2$~ps$^{-1}$, values that agree well with experiments on self-assembled quantum dots~\cite{ramsay:2010, ramsay2:2010}.

We absorb the rates in \eqref{eqn:phonondiss} into the Lindblad operators to obtain following decoherence operators:
\begin{eqnarray}
  U_+ &= \sqrt{J(N+1)}\ketbra{\psi_-}{\psi_+} \\
  U_- &= \sqrt{JN}\ketbra{\psi_+}{\psi_-}
\end{eqnarray}
where we have taken $J = J(\Lambda)$ and $N = N(\Lambda)$.

As a measure of the degree of indistinguishability of the photons produced in the emission process, we consider the HOM visibility, which is the normalised probability of same arm detections obtained over many runs of the experiment,
\begin{eqnarray}
v_{\rm HOM} = \frac{p_{\rm same} - p_{\rm diff}}{p_{\rm same} + p_{\rm diff}},
\end{eqnarray}
where $p_{\rm same}  = p(D_+ \vert D_+) + p({D_- \vert D_-})$ and $p_{\rm diff}  = p(D_+ \vert D_-) + p({D_- \vert D_+})$ with $p(D_x \vert D_y)$ being the probability of obtaining a click in detector $D_x$ conditional on the previous click having occurred in detector $D_y$.

In order to calculate $v_{\rm HOM}$ we thus need to consider photons emitted from two copies of the system $S_1$ and $S_2$, one for each input arm of the beam splitter interferometer. The joint state of the system then inhabits the space $S= S_1 \otimes S_2$.
We label the two beam-splitter input modes $a_1$ and $a_2$, and the two detection modes $c_+$ and $c_-$.
The beam splitter performs the transformation
\begin{eqnarray}
  c_+ &= \sqrt{\kappa} \frac{1}{\sqrt{2}}\left( a_1 + a_2 \right) \\
  c_- &= \sqrt{\kappa} \frac{1}{\sqrt{2}}\left( a_1 - a_2 \right)
\end{eqnarray}
where $\kappa$ is the cavity leakage rate. We assume that the field in the mode outside the cavity is directly related to the field inside, neglecting the process of escape from the cavity.
The effect on the system of a detection in the plus or minus output mode is then described by 
\begin{eqnarray}
  C_\pm &= \ketbra{g_1}{g_a}\otimes\id \pm \id\otimes\ketbra{g_1}{g_a}.
\end{eqnarray}

We could now simulate many trajectories of this system and build up an estimate of $v_{\rm HOM}$ by averaging these~\cite{carmichael:92}. Instead we use a semi-quantum master equation technique described in more detail in \ref{appA}. This allows us to find $v_{\rm HOM}$ in a single run of a master equation acting on a slightly larger Hilbert space.

\begin{figure}[htb]
  \begin{center}
  \includegraphics[width=8cm]{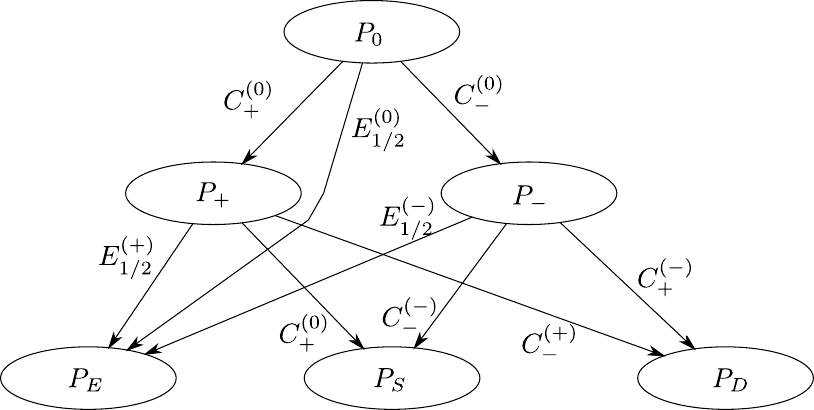}
\end{center}
  \caption{The system jump-space: On the first jump the system moves to $P_+ / P_-$ depending on the arm in which the photon is detected. After the second jump the system moves to state $P_S / P_D$ depending on whether the second photon was detected in the same or different arm to the first. At any point the system can undesirably spontaneously emit into the environment, moving to the junk state $P_E$.}
  \label{subspace_partition}
\end{figure}

When calculating $v_{\rm HOM}$ we must consider events in which photons trigger either detector, and events in which photons are spontaneously emitted into the environment. We introduce a set of process-states to record and labels these event classes (Fig.~\ref{subspace_partition}):
\begin{eqnarray}
  S_P = \left\{ P_0, P_+, P_-, P_S, P_D, P_E \right\}.
\end{eqnarray}
The process-state starts as $P_0$ and remains there until an event of interest occurs. $P_{+/-}$ represents the process-state after a single photon has been detected in the $D_{+/-}$ detectors respectively. After a second photon has been detected the process-state becomes $P_{S/D}$ depending on whether the second photon was detected in the same or different detector as the first. If at any point a photon is emitted into the environment the process moves to state $P_E$. When calculating the indistinguishability we can then ignore any population in state $P_E$, but we must include it when considering the overall efficiency of the process.

We must also identify the operators that cause the movement between the process-states. Before doing this we extend the state-space $S$ of the system to include the process-states:
\begin{eqnarray}
  S   &= S_1 \otimes S_2 \otimes S_P .
\end{eqnarray}
The detection operators are then given by
%\begin{widetext}
\begin{eqnarray}
  C_{+}^{(0)}  & = C_{+} \otimes \ketbra{P_+}{P_0}\\
  C_{+}^{(+)} & = C_{+} \otimes \ketbra{P_S}{P_+}\\
  C_{+}^{(-)} & = C_{+} \otimes \ketbra{P_D}{P_-}\\
  C_{-}^{(0)}  & = C_{-} \otimes \ketbra{P_-}{P_0}\\
  C_{-}^{(-)} & = C_{-} \otimes \ketbra{P_S}{P_-}\\
  C_{-}^{(+)} & = C_{-} \otimes \ketbra{P_D}{P_+}
\end{eqnarray}
where for example $C_{+}^{(-)}$ is the jump operator representing a second detection in the $D_+$ detector, when the first detection was in $D_-$. The spontaneous emission operators are given similarly:
\begin{eqnarray}
  E_1^{(0)} &= \ketbra{g_2}{e} \otimes \id_S   \otimes \ketbra{P_E}{P_0}\\
  E_2^{(0)} &=  \id_S  \otimes \ketbra{g_2}{e} \otimes \ketbra{P_E}{P_0}\\
  E_1^{(+)} &= \ketbra{g_2}{e} \otimes \id_S   \otimes \ketbra{P_E}{P_+}\\
  E_2^{(+)} &=  \id_S  \otimes \ketbra{g_2}{e} \otimes \ketbra{P_E}{P_+}\\
  E_1^{(-)} &= \ketbra{g_2}{e} \otimes \id_S   \otimes \ketbra{P_E}{P_-}\\
  E_2^{(-)} &=  \id_S  \otimes \ketbra{g_2}{e} \otimes \ketbra{P_E}{P_-}
\end{eqnarray}
where here $E_1^{(+)}$ represents a emission from $S_1$ acting after the first photon was detected in the $D_+$ detector.

Finally we must modify our phonon decoherence operators. The decoherence process must happen independently on each subspace, as classically separated branches of the process are unable to interfere with one another.
\begin{eqnarray}
  U_{+,1}^{(0)} &= \sqrt{J(N+1)}\ketbra{\psi_-}{\psi_+} \otimes \id_S \otimes \ketbra{P_0}{P_0}\\
  U_{-,1}^{(0)} &= \sqrt{JN}\ketbra{\psi_+}{\psi_-} \otimes \id_S \otimes \ketbra{P_0}{P_0}\\
  U_{+,1}^{(+)} &= \sqrt{J(N+1)}\ketbra{\psi_-}{\psi_+} \otimes \id_S \otimes \ketbra{P_+}{P_+}\\
  U_{-,1}^{(+)} &= \sqrt{JN}\ketbra{\psi_+}{\psi_-} \otimes \id_S \otimes \ketbra{P_+}{P_+}\\
  U_{+,1}^{(-)} &= \sqrt{J(N+1)}\ketbra{\psi_-}{\psi_+} \otimes \id_S \otimes \ketbra{P_-}{P_-}\\
  U_{-,1}^{(-)} &= \sqrt{JN}\ketbra{\psi_+}{\psi_-} \otimes \id_S \otimes \ketbra{P_-}{P_-}
\end{eqnarray}  
%\end{widetext}
with similar operators acting on the second system. We do not need decoherence operators acting on the $P_S$, $P_D$ or $P_E$, since we are only concerned with populations in, and not coherences between, these states. Moreover, we only really need to keep track of the total population in each of these subspaces - a fact that we exploit to reduce the dimension of our problem for the numerical simulations.

We form a Lindblad master equation using these $24$ Lindblad operators:
\begin{eqnarray}
  \dot{\rho} &= i\left[ \rho, H \right] + \sum_i \gamma_i \left( L_i\rho L_i^\dagger - 1/2(L_i^\dagger L_i\rho + \rho L_i^\dagger L_i) \right).
\end{eqnarray}
The $\gamma_i$ are the rates for each process. As noted earlier for the $U_{\pm, i}^{(j)}$ this rate is $1$, as the rates have been encorporated into the Lindblad operators. For the $C_\pm^{(i)}$ we need $\gamma = \kappa$, the cavity leakage rate, which we take to be $3*h$. For the $E_i^{(j)}$ we take $\gamma = 0.05ps^{-1}$, assuming a ratiative livetime of $200$~ps.

The dimension of this extended space $S$ is $4 \times 4 \times 6 = 96$. In fact we can reduce this by eliminating some unneccessary states from the subspaces. By carefully considering the basis states accessible in each the subspace corresponding to each process, we can reduce the number of system states to $9 + 6 + 6 + 1 + 1 + 1 = 24$. For our simulation we will need to calculate the density matrix for this system. As noted in~\ref{appA}, no coherences can exist between the different process-state subspaces. This reduces the number of density matrix elements we need to track to $9^2 + 6^2 + 6^2 + 1 + 1 + 1 = 116$.

\section{Results}

\begin{figure}[htb]
  \begin{center}
  \includegraphics[width=8cm]{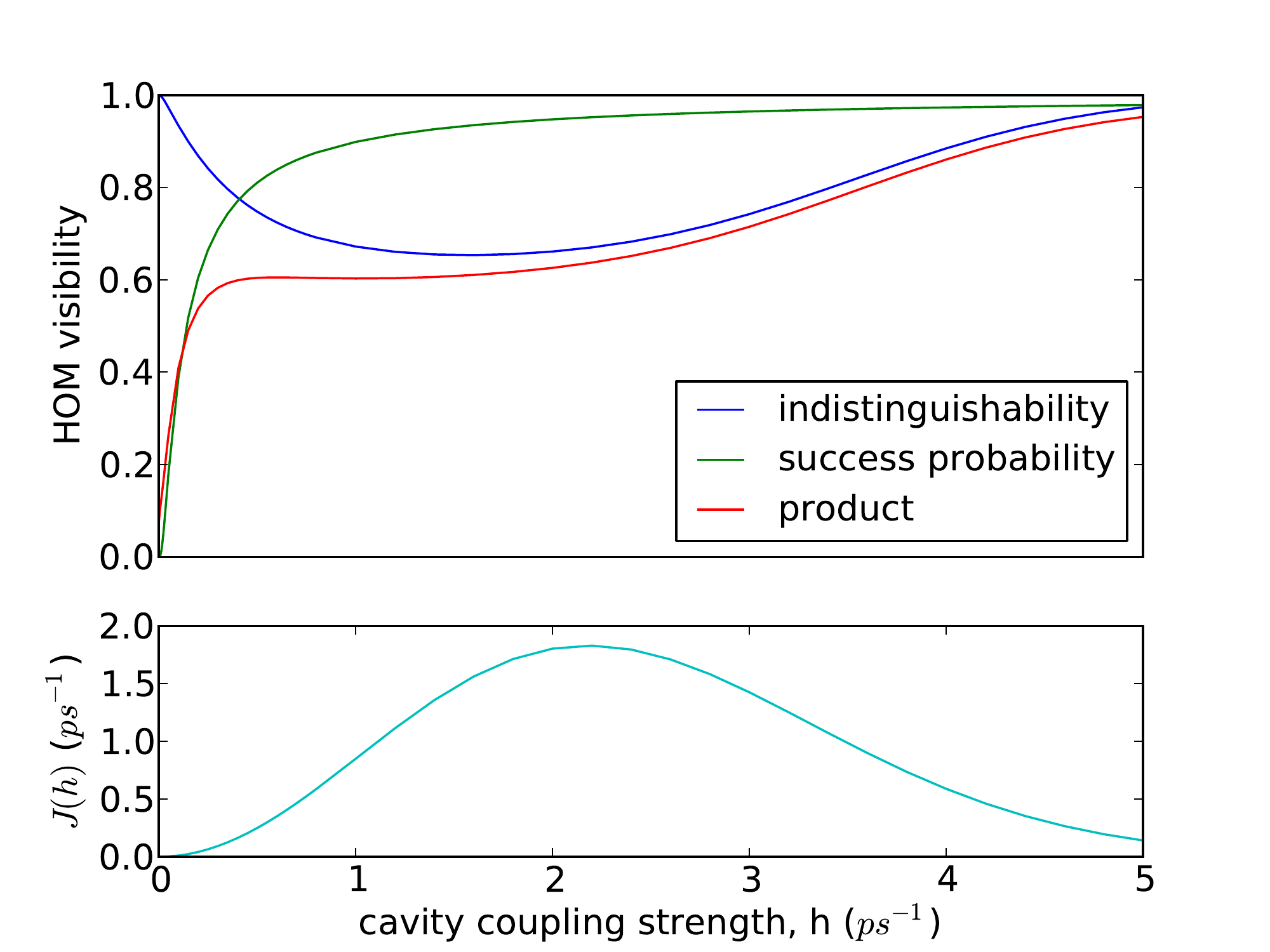}
  \end{center}
  \caption{Pulse-relax technique: calculations are performed as a function of the cavity coupling strength. Upper panel: at realistic coupling strengths ($h<2$) increased HOM indistinguishability necessarily entails a decrease in efficiency (success probability), with the product of the two (blue curve) approaching zero. Here the spontaneous emission rate is $\Gamma = 0.05~\mathrm{ps}^{-1}$. For overcoming phonon-induced decoherence and achieving a  high success probability, we must move into a region of unreasonably high $h$. % to obtain good source characteristics.
 The lower panel shows the phonon spectral density, Eq.~\ref{eq:specdens}, evaluated at the cavity coupling strength $h$, giving a rate that is directly proportional to phonon-induced dephasing during the pulse-relax process. 
 }
  \label{2LS_plot}
\end{figure}

Fig.~\ref{2LS_plot} shows the HOM visibility obtained from, and spectral density used in, simulations of the pulse-relax technique. At low coupling strengths the phonon spectral density is small, and so phonon decoherence is largely avoided giving high indistinguishability. However, in the extraction of the photon into the cavity mode is slow and thus spontaneous emission into environmental optical modes is a problem. To take account of this effect, we define the `combined HOM visibility', which is the product of success probability and bare HOM visibility, which approaches zero as $h \to 0$. Taking large coupling strengths allows one to access the region to the higher frequency side of the hump in the phonon spectral density and so could avoid this problem, but is unrealistic given the cavity parameters currently obtainable experimentally. In the experimentally feasible region ($h<1$)~\cite{vuckovic:12} we must therefore trade indistinguishability for efficiency.

\begin{figure}[htb]
  \begin{center}
  \includegraphics[width=8cm]{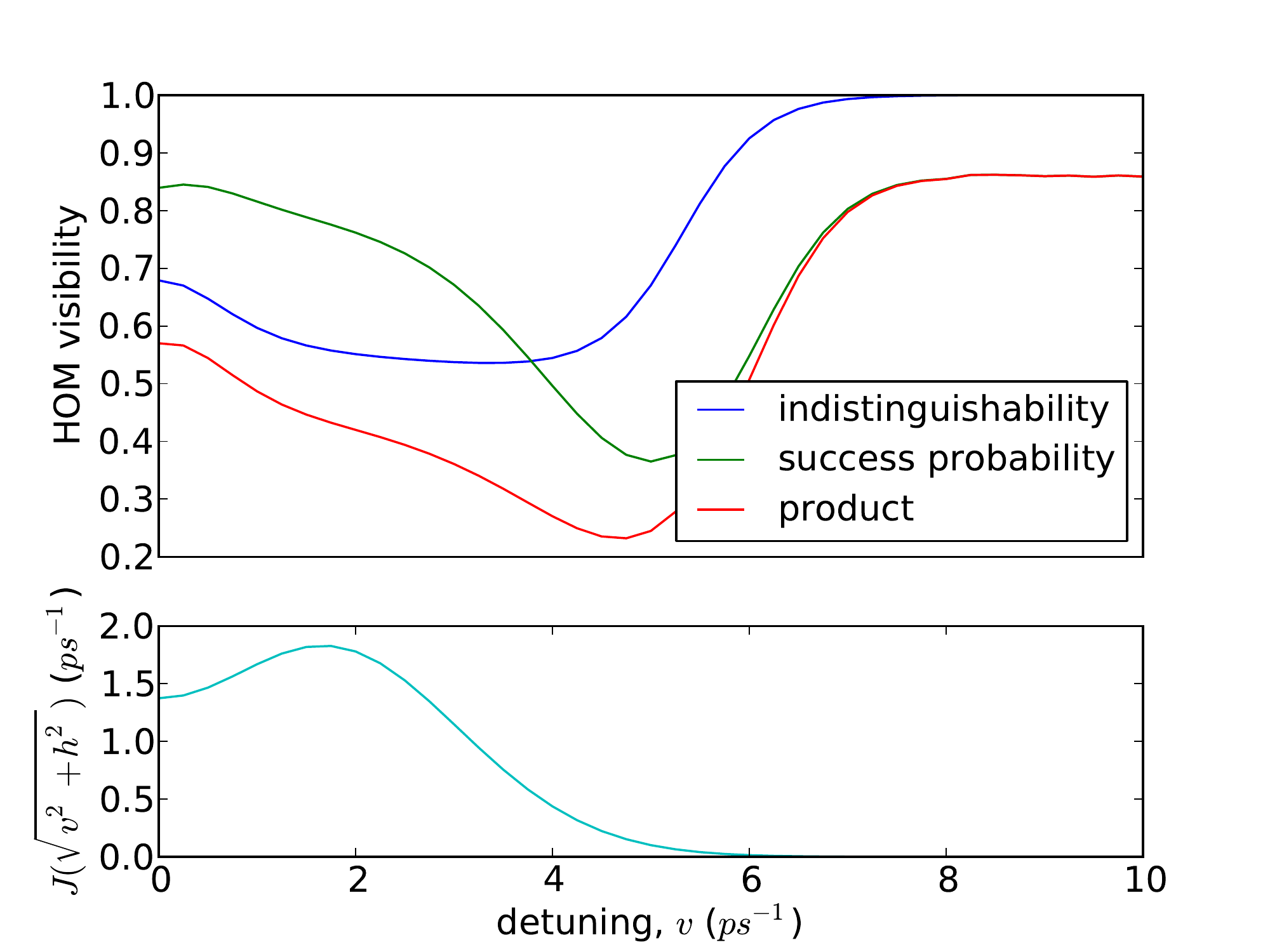}
  \end{center}
  \caption{Raman technique, $h = 0.5$: by choosing a detuning to move beyond the region of high phonon spectral density we can achieve near-perfect indistinguishability. The efficiency in this region is high enough for a feasible photon source.}
  \label{raman_plot}
\end{figure}

In contrast, the Raman procedure (Fig.~\ref{raman_plot}) avoids this trade-off. For small detuning the visibility is low, but this is because we do not get a proper Raman ground state transition unless $h \gg \nu$. If this condition is not met, the system simply undergoes a non-optimal detuned pulse-relax transition. Once we reach a detuning of around $12h = 6$ the indistinguishability and efficiency both  increase. Our choice of detuning size is limited only in that in must be small in comparison with the original energy gap between the ground and excited states. This leaves us free to use large detunings to push to frequencies above the region of high phonon spectral density. As the detuning is increased the efficiency saturates below unity. In this region both the time taken for the process and the average lifetime scale with the detuning squared, the two effects cancelling one another.

\begin{figure}[htb]
  \begin{center}
  \includegraphics[width=8cm]{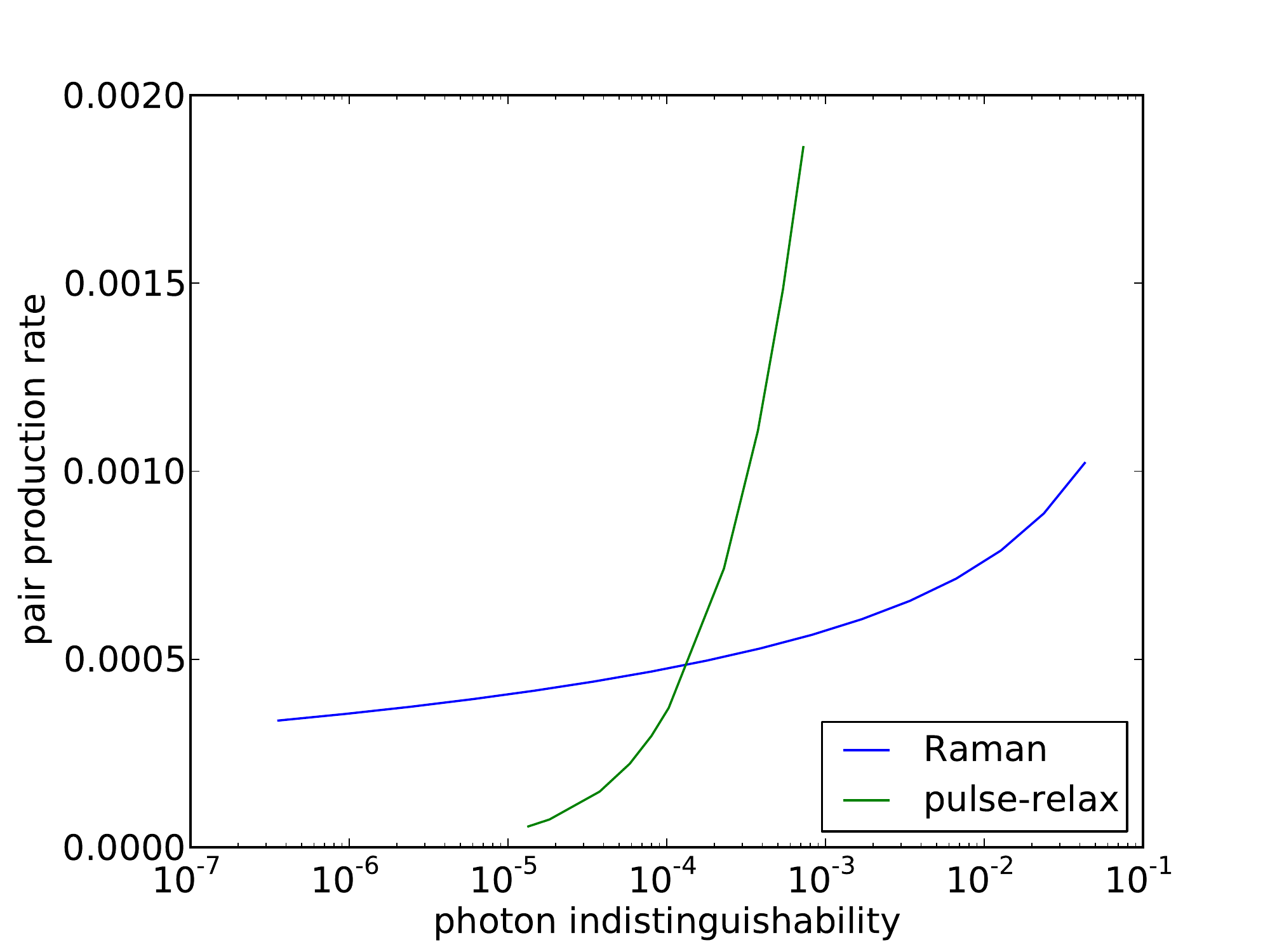}
  \includegraphics[width=8cm]{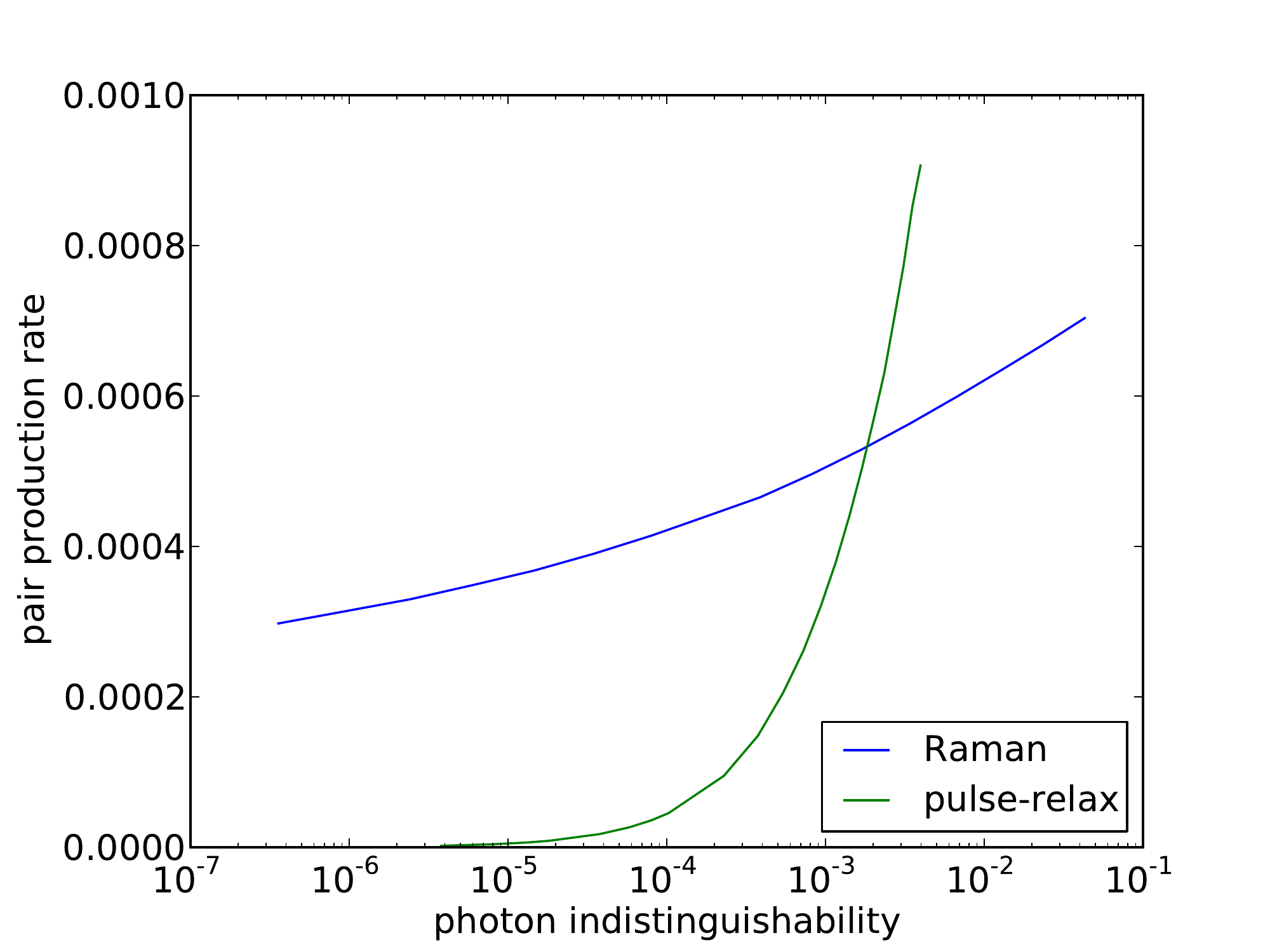}
  \end{center}
  \caption{Rate of photon pair production: with no spontaneous emission (upper), and spontaneous emission with an excited lifetime of $200$~ps (lower)}
  \label{rate_of_production_plot}
\end{figure}

In many applications where photons are used, the product of efficiency and indistinguishability may not be the most useful metric for characterising the performance of the source; often photon escape errors can be accounted for, and the indistinguishability of photons that are detected is the important figure of merit. We therefore also calculate the rate of production of pairs of photons of a given indistinguishability using each approach (Fig.~\ref{rate_of_production_plot}). For our production rate we take
\begin{eqnarray}
  r_f = \frac{e_f}{t_f}
\end{eqnarray}
where $e_f$ is the efficiency and $t_f$ is the time taken for $99\%$ of the runs to have completed (possibly unsuccessfully), for parameters chosen to obtain a given indistinguishability, $f$. This figure is somewhat approximate as it takes no account of how successful runs (where two photons are emitted into the correct modes) and unsuccessful runs are distributed within the process run time, and no allowance is made for time taken to reset the system in the event of a failure. The effect of the former is minor since in our model spontaneous emission can occur uniformly at any point of the process. We will revisit the effect of the latter shortly.

Even in the absence of spontaneous emission (Fig.~\ref{rate_of_production_plot}, upper panel), the Raman procedure is quicker than the pulse-relax process at generating photons of a sufficiently high level of indistinguishability. With the dephasing parameters chosen in our model this occurs for indistinguishability of greater than $99.99\%$. In our model, spontaneous emission is the only process degrading the efficiency -- without it we have perfect efficiency and so neither of the potential shortcomings discussed in the previous paragraph apply.

When spontaneous emission is added, we see a similar pattern but the indistinguishability threshold is lower ($99.9\%$ working with a spontaneous emission rate of $0.05$~ps$^{-1}$). This is an upper bound, as here the reset time becomes important. The efficiency of the Raman procedure remains fixed at about $80\%$, requiring on average $1.25$ runs per pair. In contrast, the efficiency of the pulse-relax procedure heads towards zero, meaning that many attempts will be needed to produce a pair. If the time taken to reset the system (to the excited state $\ket{e}$ in which we have assumed the pulse-relax system starts) is large, the Raman procedure will become advantageous at a far lower threshold.

\section{Conclusion}

To conclude, we have developed a realistic and microscopically justified model of the impact of phonons on solid state single photon sources. We used a modified, `semi-quantum', master equation method for the efficient calculation of coincidence rates, without having to resort to a quantum Monte-Carlo simulation approach.  In systems where phonon dephasing and spontaneous emission are the dominant loss channels, and with current limits on the obtainable cavity coupling $h$, we find that a Raman technique is preferable to the pulse-relax approach for producing highly indistinguishable photons that would be suitable for the most demanding applications of quantum information processing. %\include{braket}

\section{Acknowledgements} We thank Ahsan Nazir, Simon Benjamin, Pieter Kok and Sean Barrett for useful discussions. This work was supported by the Engineering \& Physical Sciences Research Council (UK) and the National Research Foundation and Ministry of Education, Singapore. B. W. L. thanks the Royal Society for a University Research Fellowship. 

\appendix
\section{Semi-quantum Master Equations}
\label{appA}

We are often interested in `observable events' in quantum systems, such as the emission of a photon. When an event is observed the system undergoes a transition due to wave function collapse. During a period when no event is observed the system evolves according to a conditional master equation, reflecting the fact if events could be observed but are not, then this also informs our knowledge of the state. In order to answer questions about the probabilities and time distributions of events or chains of events, one approach is to use a quantum jump master equation to generate individual trajectories of the system. In each timestep we decide probabilistically whether an event should occur. If it does occur then the system collapses according to a quantum jump; if it does not occur then the system evolves conditioned on no jump occurring. Statistics about the quantities of interest are built up as many trajectories are created.

Here we look at a different approach to calculating the quantities relating to events that occur in such systems. Instead of simulating multiple trajectories of the system, we efficiently increase the size of the statespace to record the information of interest. This allows us to calculate the desired system properties, and their time evolution, by solving a single master equation. 

\begin{figure}[htb]
  \begin{center}
  \includegraphics[]{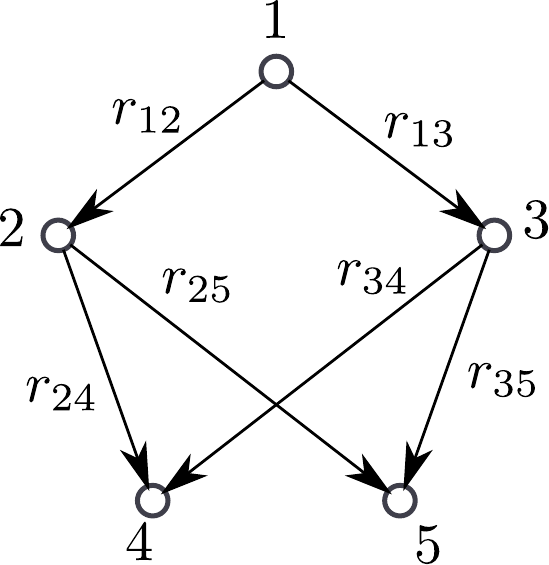}
  \end{center}
  \caption{A classical continuous-time Markov chain, with statespace $W = \{1,2,3,4,5\}$. The edge weight, $r_{ij}$, represents the transition rate from state $i$ to state $j$. If $\rho_i(t)$ is the population in state $i$ at time $t$, the system is governed by the rate equations $\dot{\rho_i} = \sum_{j\in W} r_{ji}\rho_j$.}
  \label{markov_chain}
\end{figure}

Consider a system that can exist in a number of different states. Let a movement between these states constitute an event, and assume that each kind of event happens at a given rate. Such a system is heavily reminiscent of a classical continuous time Markov chain (CTMC), which can be represented as a graph with the states as nodes and the edges events weighted by the transition rates (Fig.~\ref{markov_chain}). 
Given an initial state $i$ in a chain of size $n$, we can calculate the probability that at a later time $t$ the chain is in state $j$, by solving the rate equations - a set of $n$ ordinary differential equations.

Quantum systems differ from classical continuous-time Markov chains due to quantum superposition. We are not able to simply record the population in each quantum state as inter-state coherences are also important. Using a Markov chain to model the whole quantum system is not possible by definition - systems that can be modelled in this way do not exhibit quantum behaviour.

In what follows it is helpful to explain carefully what we mean by `state'. In quantum systems a state is usually a vector in the Hilbert space of the system. We shall call this a quantum-state. We can also refer to the `state' of the overall process, considering for example a system that has emitted a photon to be in a different process-state to one which has not. A system changes process-state when an event is observed.

Whereas thinking in terms of the CTMC is not useful when considering quantum-states, it is an effective way to think about process-states. As process-states are separated by an observed event, no coherences can exist between the two histories, making a CTMC approach feasible. Of course, process-states alone are not enough to model the whole system. Each process-state needs its own copy of the system attached to it. We can think of a Markov chain with a copy of our system at each node, where the transition rates are determined by the jump Lindblad operators corresponding to the events. Another way of thinking about this is that we extend our overall space with a set of process-states, to allow us to record events in the system.

Formally, we take a set of process-states $S_P$, transitions between which correspond to our observable jump events described by jump operators $J_Q^{(i)}$. We extend the Hilbert space of our quantum system $S_Q$ by forming the tensor product:
\begin{eqnarray}
  S = S_Q \otimes S_P.
\end{eqnarray}
The new Hamiltonian is given by
\begin{eqnarray}
  H = H_Q \otimes \id.
\end{eqnarray}
If event $J_Q^{(i)}$ causes a transition from system state $a$ to $b$ we say it is of type $(a, b)$. Its action on the extended system $S$ is described by
\begin{eqnarray}
J^{(i)} = J_Q^{(i)} \otimes \ketbra{b}{a}.
\end{eqnarray}
For any other Lindblad operators acting on the system $L_Q^{(i)}$, we need to create a set of size $|S_P|$ Lindblad operators - one to operate on each subspace independently:
\begin{eqnarray}
  s(L_Q^{(i)}) = \left\{ L_Q^{(i)} \otimes \ketbra{j}{j}, j \in S_P \right\}.
\end{eqnarray}

At first glance it might appear that we have increased a system of size $m = |S|$ to size $mn$, where $n = |S_P|$. Whilst this is true, the situation is not as bad as it seems at first, because the coherences between the different subsystems are unimportant -- instead of a density matrix of size $(nm)^2$ we can use a system of size $nm^2$, an increase linear in the number of system states. In practice we can often do better than this by eliminating unnecessary states from some of the subsystems.  

As a simple concrete example, consider a resonantly driven two-level system $S_Q = \{ \ket{g}, \ket{e} \}$, with Hamiltonian
\begin{eqnarray}
  H_Q = \Omega \left( \ketbra{g}{e} + \ketbra{e}{g} \right).
\end{eqnarray}
Suppose that it is also possible for the system to spontaneously emit from state $\ket{e}$ into the environment - a transition described by the jump operator
\begin{eqnarray}
  J_Q = \ketbra{g}{e}.
\end{eqnarray}

We are interested in knowing about the time distribution of the first time a photon is emitted. We add the two process-states $0$ and $1$, indicating whether the event has occurred or not. Our new system is described by:
\begin{eqnarray}
  S =& \left\{ \ket{g0}, \ket{e0}, \ket{g1}, \ket{e1} \right\} \\
  H =& \Omega \left( \ketbra{g0}{e0} + \ketbra{e0}{g0} + \ketbra{g1}{e1} + \ketbra{e1}{g1}  \right) \\
  J_1 =&  \ketbra{g1}{e0} \\
  J_2 =& \ketbra{g1}{e1}.
\end{eqnarray}
The population in the $1$ subspace at time $t$ will give us the probability a photon has been emitted by this time.

If we only cared about this distribution, we could reduce the size of the system by eliminating the state $\ket{e1}$, removing the last two terms from the Hamiltonian as well as $J_2$.
\\

\end{document}